# Dual-aptamer Drift Cancelling Techniques to Improve Long-term Stability of Real-Time Structure-Switching Aptasensors

Ya-Chen Tsai[1], Wei-Yang Weng[2], Yu-Tong Yeh[1], Jun-Chau Chien[1,2]

[1]Department of Electrical Engineering

[2]Graduate Institute of Electronics Engineering

National Taiwan University

*Abstract*

This paper presents a dual-aptamer scheme to cancel the signal drifts from structure-switching aptamers during long-term monitoring. Electrochemical aptamer-based (E-AB) biosensors recently demonstrated their great potential for *in vivo* continuous monitoring. Nevertheless, the detection accuracy is often limited by the signaling drifts. Conventionally, these drifts are removed by the kinetic differential measurements (KDM) when coupled with square-wave voltammetry. Yet we discover that KDM does not apply to every aptamer as the responses at different SWV frequencies heavily depend on its structure-switching characteristics and the redox reporters' electron transfer (ET) kinetics. To this end, we present a "dual-aptamer" scheme that uses two aptamers responding differentially to the same molecular target for drift cancellation. We identify these paired aptamers through (1) screening from the existing aptamers pool and (2) engineering the signaling behavior of the redox reporters. We demonstrate their differential signaling to ampicillin and ATP molecules and show that the aptamer pair bears common drifts in undilute goat serum. Through cancellation, sensor drift is reduced by 370-fold. Benefiting from the "differential" signaling, the recording throughput is also doubled using differential readout electronics. The authors believe the proposed technique is beneficial for long-term *in vivo* monitoring.

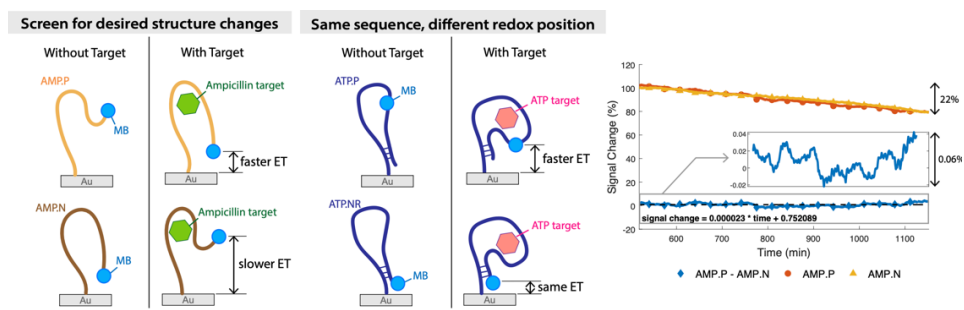

## *Introduction*

Electrochemical aptamer-based (E-AB) sensors are redox-labeled synthetic nucleic acids that can perform specific monitoring of bio-molecules in complex samples in real-time continuously [1-10]. They are often engineered into molecular "switches" that undergo structure switching upon target binding, and the responses are detected without needing additional reagents. Several recent works have demonstrated its use in long-term *in vivo* monitoring, opening up new applications such as personalized drug dosing. These E-AB sensors can further couple with miniaturized integrated circuits (ICs) to form wearable or implantable electronics that suits non-ambulatory use [24].

Nonetheless, several challenges remain when applying E-AB sensors for long-term monitoring. These include aptamer desorption from the electrode surfaces, aging of the reporting molecules and the aptamers, the nature and non-nature degradation of the self-assembled monolayer (SAM), biofouling, and enzymatic attacks from the nucleases [13,14]. All of these can cause unwanted baseline drifts and varying sensitivity, thus increasing the readout inaccuracy. Techniques that can reduce or even mitigate these impacts are highly desirable.

The most widely used drift-mitigating technique in the E-AB sensors is the kinetic differential measurement (KDM) [3, 4, 7, 8, 17]. It measures the varying electron transfer (ET) kinetics from the redox reporter at two square-wave voltammetry (SWV) frequencies, each showing "signal-on" and "signal-off" responses, and applies difference operation between the two consecutive samples to remove drifts that appear common in both measurements. The inversion in the sensitivity arises from the changing sampling instants in the electron transfer process, as shown in Figure 1, and is crucial to avoid canceling the signals. KDM has been applied in several recently reported E-AB sensors, including doxorubicin [3], vancomycin [17], and ochratoxin [51]. Nonetheless, we also observe some E-AB sensors do not signal in the same way. For example, the irinotecan aptamer presented in [7] does not have two frequencies of inverting sensitivity. To overcome this, the authors co-immobilize a non-responsive strand on the same electrode to create an intentional reference measurement. On the other hand, the ampicillin aptamers in [33-34] exhibit both signal-on and signal-off responses when measured in buffer, but these are not observable in our animal study in rat whole blood [24]. Whether an aptamer poses such a sensitivity inversion is highly dependent on how the electrons from the redox reporters are transferred, and factors such as aptamer rigidity, diffusion distance, over-potentials, and the percentage of redox reporters that undergo complete redox cycles can all play an impactful role (Figure S1). It is therefore likely that KDM could be inapplicable in some of the structure-switching aptamers. Another drawback in KDM is the extended per-sample acquisition time (> 10 seconds) due to the need for multi-frequency measurements. This is generally not an issue in applications such as therapeutic drug monitoring (TDM) but could become a limiting factor in applications that require higher temporal resolution, such as the study of neurotransmitter release in the brain [48].

In addition to KDM, several other techniques are proposed to reduce the signaling drifts. Notably, [16] employs a dual-labeling approach that conjugates methylene blue (MB) and anthraquinone (AQ) at different locations along the aptamer sequence. By having one label closer to the "root" of the aptamer, its redox current will remain insensitive to the changing conformation and can serve as the reference in a ratiometric measurement. Nonetheless, this requires two redox labels and a wider scanning potential range, which can severely impact the stability of the aptamer signals as highlighted in [13]. Different electrochemical methods can be applied to improve signaling stability. For example, [49] employs cyclic voltammetry (CV) and extracts concentration information from the voltage gap between the reduction and the oxidation potentials, [26] uses electrochemical impedance spectroscopy (EIS) to extract the charge-transfer resistance, and [25] applies chronoamperometry (CA) to probe the electron transfer kinetics directly. Other methods to improve the E-AB sensor's long-term stability includes the use of hydrogel protecting [37] and cell-membrane-mimicking phosphatidylcholine-terminated monolayers [23]. Both are designed to avoid non-specific adsorption of bio-molecules and cells.

A dual-aptamer scheme is proposed in this paper to improve the long-term stability of the E-AB sensors. Our idea is to utilize a pair of redox-labeled aptamers with identical or similar sequences that can respond differentially to the target molecules while exhibiting common drifts. By subtracting the signals from one another, drifts common to both aptamers are removed while preserving and amplifying the desired sensor signals. The most crucial aspect of this dual-aptamer scheme lies in developing these aptamer pairs. Two methods are presented in this paper. In the first approach, we screen the aptamers targeting a similar group of molecules. Specifically, we focus on those selected for small-molecule antibiotic drugs [3][33] and identify the pairing of the ampicillin and doxorubicin aptamers (Figure 1(a)). The second method is by modifying the redox position on constructs of the same sequence. With the presence of the targets, both aptamers will change their conformations similarly but the ET kinetics of the redox reporters are modulated differentially. We demonstrate this idea using the ATP aptamers developed in [6], which integrate a linker and a displacement strand with the intrinsic ATP-binding aptamers. In the first construct, we conjugate the redox reporters at the 3' end such that their diffusion distance increases upon target binding. This lowers the ET rate and leads to current reduction. In the second construct, we position the redox reporter internally next to the linker. The binding of ATP molecules then promotes the formation of the hairpin structure which drags the reporter molecule toward the underlying electrodes. This shortens the diffusion distance and increases the ET rate as well as the readout currents. Both conformations are illustrated in Figure 1(b). Our 24hr measurements in undilute goat serum show that the drifts can be reduced by 370-fold. We also study the impact of changing pH, ion concentrations, and temperature on the paired aptamer responses and emphasize the need for signal compensation using multi-parametric sensor fusion. The sample acquisition rate can also benefit from the dual-aptamer scheme by using a custom two-channel readout electronics monitoring both electrodes simultaneously. Applications requiring a higher temporal resolution, such as detecting neurotransmitters in the brain [47, 50], can thus benefit from enhanced throughput.

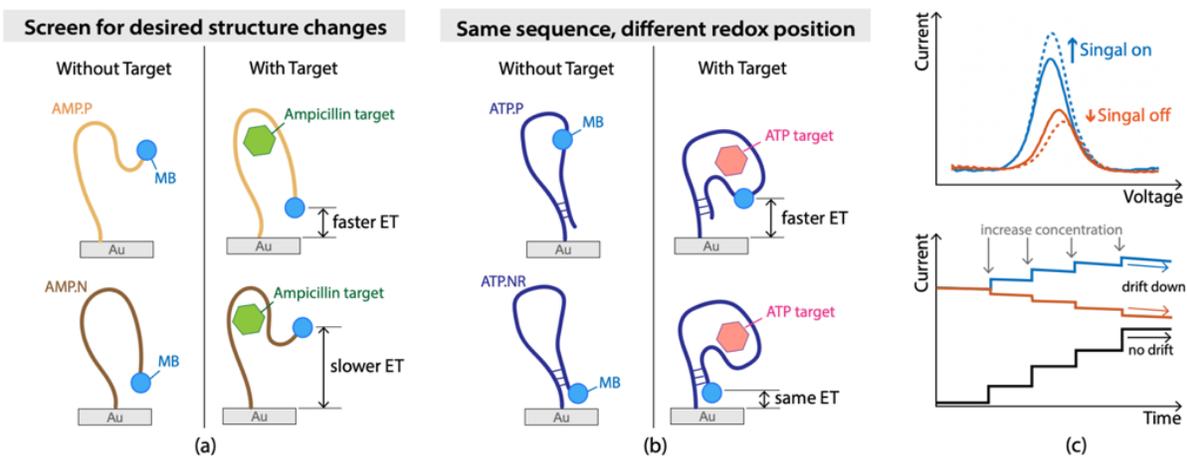

Figure 1. Two approaches are utilized to develop the aptamer pairs. (a) In the first approach, we screen for aptamers that exhibit opposite responses to the same molecular target. This is demonstrated by detecting ampicillin, an antibiotic drug, using ampicillin (AMP.P) and doxorubicin (AMP.N) aptamers. (b) In the second approach, we change the position of the redox label on the aptamer to formulate differential signaling upon target binding and demonstrate this concept using an ATP-binding aptamer. (c) The illustrations of the voltammograms and the time-domain traces from the paired sequences with the presence of the targets. The drifts common to both sequences are canceled after signal subtraction.

## *Materials and Methods*

**Sensor Devices.** We prepare the sensor devices as wiring probes to emulate an implantable device for real-time sensing in the vein [4]. To this end, we employ 250-μm diameter gold (Au) wires (Thermo Fisher Scientific Inc., MA, USA) as the working electrodes (WE) and define the aptamer sensing region (~0.25 mm$^2$) using heat shrink micro-tubings (McMaster-Carr Inc.). We utilize silver/silver-chloride (Ag/AgCl) and platinum (Pt) as pseudo reference (RE) and counter electrodes (CE), also in the form of wires (A-M Systems Inc., Carlsborg, WA). Figure S2 depicts a photo of the sensor device. Prior to aptamer immobilization, the WEs are cleaned in solvents (to remove organics particles) and two rounds of electrochemical cleaning using 200 mM and 50 mM $H_2SO_4$ with cyclic voltammetry (CV). The CV potentials are scanned between -0.3 and 1.5 V at scan rates of 1 and 0.1 V/sec, respectively. After cleaning, the electrodes are stored in de-ionized water.

**Aptamer Functionalization.** DNA aptamers and ATP solutions are purchased from Genomics (New Taipei City, Taiwan) and Thermo Fischer Scientific (MA, USA), respectively. All other chemicals are purchased from Sigma Aldrich Inc. (TPE, TWN). Upon receiving, the aptamers are prepared at 100 μM with de-ionized water before use. To immobilize the aptamers onto gold WEs, the disulfide bonds on the 5' ends of the DNA are first broken by mixing and incubating 1 μL of aptamer solution with 2 μL Tris (2-carboxyethyl) phosphine (TCEP) at 28.7 mg/mL for 40 minutes in dark. The 3-μL solution is then diluted with 97 μL 1× saline sodium citrate (SSC) buffer, reaching a final aptamer concentration of 1 μM. This concentration is selected to achieve the desired aptamer surface density (~1%). Next, the cleaned Au electrodes are incubated in the aptamer solution for 1 hr, and the surface is further passivated with 10 mM 6-mercapto-1-hexanol (MCH) for 2 hr. Both steps are performed at room temperature. The devices are stored in a 1× SSC buffer at 4°C overnight to allow the stabilization of the self-assembly monolayers. The aptamer sequences are provided in Table S1.

**Electrochemical measurements.** We employ square-wave voltammetry (SWV) using a commercial potentiostat (Palmsens Inc., GA, Houten) for reading the aptamer states. The potentials are scanned from -0.55 to -0.15 V at a 36-mV square-wave amplitude. Both the SWV frequencies and the step sizes are adjusted to ensure each data acquisition falls within five seconds. To facilitate multiplexed detection, the potentiostat interfaces a low-leakage-current 1-to-8 switching network integrated circuit (ADG728 from Analog Device Inc., CA, USA) before connecting to the sensing electrodes. We employ a Faraday cage to minimize environmental interferences and perform all measurements at room temperature. The responses of the aptamers against different target molecules are tested in 1× SSC buffer. Prior to target challenging, SWV is performed repeatedly for 40 minutes continuously on newly prepared devices to ensure a stable baseline. Responses at each new target concentration are also measured for at least 15 minutes to ensure equilibrium. To study the efficacy of the dual-aptamer-based drift cancellation technique, we modulate the buffer's ionic strength and pH using 20× SSC, 1 M sodium hydroxide (NaOH), or 200 mM sulfuric acids ($H_2SO_4$). Solution pH is further calibrated using a pH meter (HANNA Inc., RI, USA). To characterize the long-term performance of the aptamer sensors in serum, we implement a continuous-flow system using a peristaltic pump (Dogger Inc., NWT, TWN) as a simulation of blood circulation. The aptamer-functionalized sensing probes are introduced through 14G catheters. The data acquisition is performed automatically through MATLAB controls.

**Data analysis.** All data analyses were performed by MATLAB scripts coded in-house. Although it has been reported that signal amplitudes from SWV fluctuate from sensor to sensor, we assume the effect of sensor variation would be negligible after normalization to baseline values. We extracted the peak currents from SWV as the signals and did two analyses after normalization: (1) we plotted the signals as a function of time and made use of the results to perform drift cancellation. Although all the working electrodes were impacted by drift, due to different affinities, surface defects, and so on, they may suffer from different scales of drift. To maximally reduce the noise signals from drift, we did a linear approximation to the baseline data of each working electrode, then we scaled the whole data to make the two slopes identical and subtracted the two groups of scaled data. As a result, it was expected to have a zero slope throughout the baseline period in the subtraction results (Figure S6). This analysis was applied to all data processing throughout this work, including data from titration, modulation, and circulation.

Unfortunately, the scaling constants were not repeatable among different experiments. The reasons are speculated as different surface areas and surface defects that cause different drifts between each experiment. However, the maximum time delay from curve fitting and constant extraction in our experiments is only 8 ms, which should be acceptable in clinical use.

Also (2) we plotted the signals as a function of the target concentration. From the Langmuir adsorption model, we approximated the function as a Langmuir curve and extracted the concentration where the signal equaled half of the saturation level as the dissociation constant ($K_D$) of the DDC system. We use this analysis to verify the effectiveness of the titration experiments.

### *Results*

Figure 2 demonstrates the signal-on and signal-off responses from the paired aptamers measured in the buffer. In the ampicillin detection, the AMP.P aptamer [33-34] and the AMP.N (originally selected for detecting doxorubicin [3]) exhibit opposite polarities in their response curves yet achieve a similar dissociation constant ($K_D$) of 2.75mM $\pm$ 1.15mM and 2.76mM$\pm$1.04mM, respectively. After subtracting one from the other, the magnitude of the responses is doubled from 12% to 26% at 2.44 mM at 100Hz, and the signal-to-noise ratio (SNR) is enhanced by a factor of 52 %. Here we define SNR as the ratio between the signal change at $K_D$ concentration and the limit-of-detection (LoD = $3\sigma$, where $\sigma$ is the standard deviation from the measured time-domain waveform without any target). The measured "composite" $K_D$ is 3.19$\pm$0.93 mM at the SWV frequency of 200 Hz, which closely matches our estimation from the individual $K_D$. We further perform repeated tests, and the ampicillin aptamer pair achieves a worst-case inaccuracy of 14.9% within the clinical range (71.5uM ~ 5.7mM [36], Figure S7(a)). On the other hand, the ATP aptamer pairs, which differ only by the conjugation positions of the redox label, show a "signal-on" and a "non-responsive" behavior. By conjugating the methylene blue internally in the ATP.P construct, we observe significant changes in the measured SWV current (238% at 2.7mM). This is expected as the intrinsic ATP aptamer forms a hairpin structure upon binding to the ATP molecules [6] and modulates the MB diffusion distance relative to the Au electrode. On the other hand, when the MB is conjugated at the 3'-end, minute changes other than signal drifts are observed. In other words, the conformation change from binding to the ATP molecules does not cause sufficient modulation in the MB electron-transfer (ET) kinetics. Yet, it is this "non-responsive" behavior that facilitates the canceling of the drifts in the long-term monitoring, as will be demonstrated shortly. After applying the difference to the two responses, we achieve a composite $K_D$ of 7.41 mM±0.16 mM. Note that this affinity is lower than the initially reported construct designed for fluorescence detection [6] We believe this is due to the differences in the aptamer formats (freely floating versus being immobilized onto the electrodes). Nonetheless, repeated measurements show 18 % inaccuracy in the clinical range (1.5mM~3.15mM [21, 35], Figure S7(b)) and fall within the acceptable inaccuracy (20%) for most of the clinical applications [21, 35]. We believe the affinity can be further improved by re-engineering the configurations and the sequences of the linkers and the displacement strands, as those in [6]. This is beyond the scope of the paper and will be our future work.

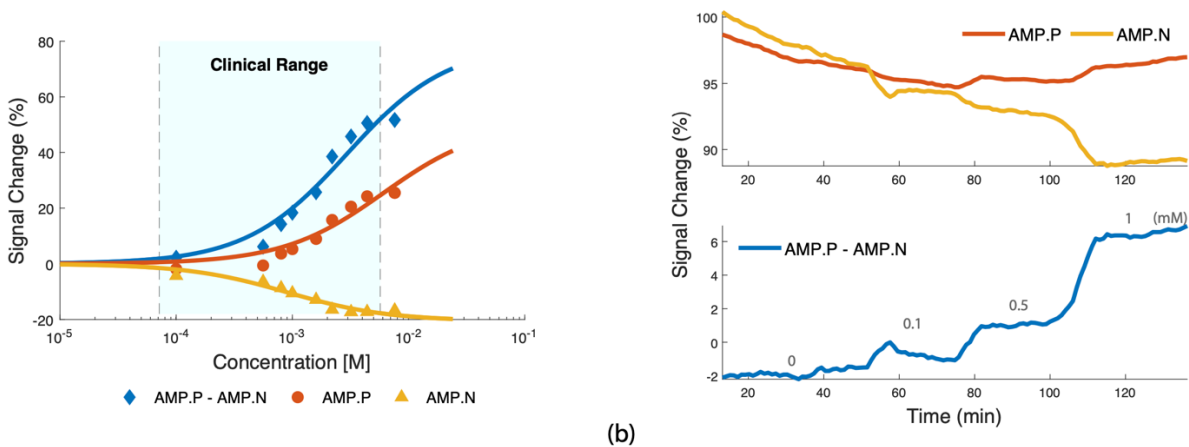

Figure 2. (a) The measured titration curves and (b) time-domain traces of the ampicillin aptamer pair. Benefiting from their differential responses, the combined signal is boosted by two folds, leading to better detection signal-to-noise ratio (SNR). For example, the AMP.P aptamer exhibits a 12% signal change at ~2.4mM, whereas the AMP.N aptamer shows a -14% change. The combined signal change is 26% after subtraction.

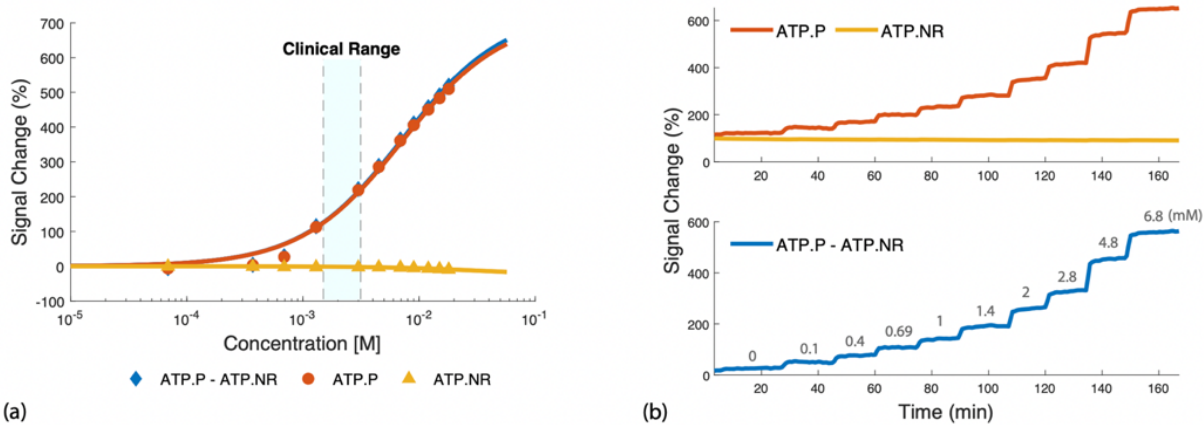

Figure 3. (a) The measured titration curves and (b) time-domain traces of the ATP aptamer pair. As the ATP concentration increased up to 2.7 mM, the ATP.P aptamer shows a 234% signal change and the AMP.NR aptamer exhibits a -4.5% change. The combined signal change is 238% after subtraction. Due to the drastic differences in their sensitivity, we term the second aptamer non-responsive (NR).

.

We hypothesize that two aptamers of identical or similar sequences will exhibit a similar trend in their signaling drifts. If the two strands respond differently to the presence of the targets, one can remove the unwanted drifts by subtracting one data from the other while preserving the desired sensing signals. To study this, we set up a continuous flow experiment using a peristaltic pump and insert the aptamer-functionalized sensing electrodes into the tubing through catheters (Figure S3). The wire electrodes are prepared similarly to those presented in [37], and the two working electrodes (WE) are functionalized with two different but paired aptamers. We conduct SWV measurements on each electrode every two minutes at a flow rate of 4.5 mL/min. This is close to the maximum flow rate in the human veins. To avoid the pump motor from interfering with our redox currents, we shield our potentiostat (Palmsens Inc.) using a Faraday cage. Figure 4(a) shows the long-term recording results in undilute goat serum from the ampicillin aptamer pair. We first normalize the signals from each channel to its corresponding starting point to remove the variation from the electrode area and quantify the rate of drifts in each channel (%/minute) from the first ten minutes of recording. Next, we scale all the succeeding data from the second channel by the ratio of the two drift rates and apply a differential operation to remove the drifts (Figure S6). Our measurement results show that the drift rates in each channel are -0.037 and -0.032 %/minute, respectively, and are reduced by 1600 folds to 0.000023 %/minute over the course of 12 hours.

Figure 2(b) further shows the measured time traces with changing ampicillin concentration, demonstrating that the real-time signals are preserved.

Figure 4(b) shows a similar long-term recording of the ATP aptamer pair. Here, the scaling factor is found to be unity as both channels show a nearly identical drift rate of -0.045 %/minute in the first ten minutes of recording. Interestingly, the drift rates in each channel can vary from time to time, as indicated by the different signaling slopes. This phenomenon matches closer to the actual monitoring *in vivo* [13]. Similarly, our dual-aptamer drift-cancellation scheme mains effective and can reduce the drift rate by 3000 folds, achieving a final drifting of -0.000015 %/minute.

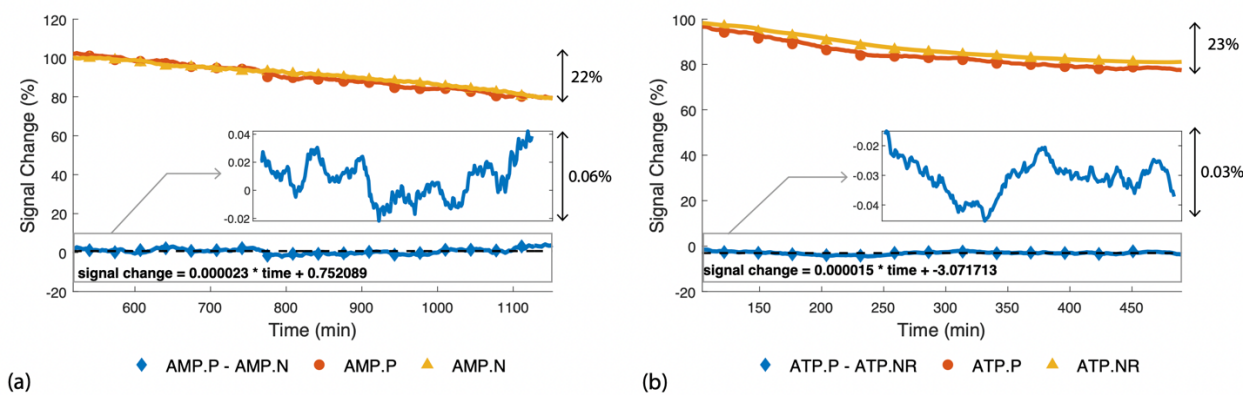

Figure 4. (a) The long-term measurements of AMP.P/AMP.N pair in the circulation system. Both AMP.P and AMP.N aptamers reduce their SWV currents by 22% over the course of 10 hours. Such a fluctuation is reduced to 0.06% after cancellation. The final drift rate is 0.000023 (%/ minute). (b) The long-term measurements of AMP.P/AMP.N pair. The signals from ATP.P and ATP.NR are both reduced by 23% over the course of 8 hours. Such a fluctuation is reduced to 0.03% after cancellation. The final drift rate is 0.000015 (%/ minute). Both experiments are measured in undilute goat serum.

Various factors, including ionic concentrations, pH, and temperature, can impact aptamer signaling and drifting behavior. Generally, we assume these parameters are relatively constant in the *in vivo* environment. However, certain diseases, such as sepsis, can incur significant alternations in these parameters [38-39]. For example, in critically-ill sepsis patients, body temperature and pH can vary from 37°C to 38.3°C [39] and 7.4 to 6.9, respectively [38]. Therefore, a detailed understanding of their impacts on drift-tracking behavior in the dual-aptamer system is mandatory. To this end, we will employ the ampicillin and ATP aptamer pairs as our study vehicles. We aim to understand how these aptamer pairs respond and whether the drift cancellation is still applicable, and if not, what are the potential solutions.

First, we study the impact of pH levels on aptamer responses. Aptamers generally respond to pH changes and can be configured as pH sensors [42-45]. From our measurements shown in Figure 5(a), we indeed observe signal changes when buffer pH is varied between 4.8 and 8.8. On the other hand, Figure 5(a) also shows that the drift cancellation is still effective, and the drift rate is reduced by > 12 folds. If needed, a pH sensor can be further included to compensate for the responses of each aptamer. Next, we modulate the ionic strength and the concentration of $Mg^{2+}$ in the buffer. As shown in Figure 5(b) and Figure 5(c), the aptamers also exhibit different sensitivities to these two parameters, and their influence cannot be removed directly through simple subtraction. The net response is negative with increasing ionic strength, whereas it turns positive with increasing $Mg^{2+}$ concentration. As such, it is suggested to include multiple ion sensors, such as ion selective membranes [40], to track those variations for signal corrections. We also modulate the medium temperature using a thermal controller and perform recording at room temperature, 37°C, and 45°C (setup shown in Figure S3). Figure 5(d) and Figure S8(d) demonstrate that the two aptamers respond drastically different to the temperature changes. The temperature sensitivities differ by 1.5x and 2.6x in the ampicillin and ATP aptamer pairs, respectively. We thus conclude that the dual aptamer drift cancellation scheme cannot mitigate the impact of temperature fluctuation. To this end, a dedicated temperature sensor is needed to compensate for the temperature impact. Figure 5(d) demonstrates the compensated results. Here we first measure and model the temperature sensitivities for each aptamer. Next, we apply signal correction in each channel, followed by drift cancellation. The temperature impacts are mostly removed and the residual spikes arise from processing artifacts. It is worth mentioning that highly accurate and low-power temperature and pH sensors can now be implemented using VLSI semiconductor technologies as part of a miniaturized multi-parametric sensor fusion system [41, 52, 53].

The dual aptamer scheme has the potential to double the measurement throughputs. Unlike KDM where two SWV measurements need to be performed sequentially on the same working electrode, the two aptamers can be interrogated simultaneously with two readout electronics. Figure 6 illustrates the block diagram, and a setup photo is shown in Figure S9. We employ two trans-impedance amplifiers (TIA) to interface with the two working electrodes, and their differences are amplified by an instrumentation amplifier (IA). The signals are then digitized with an analog-to-digital converter (ADC), and "differential" voltammograms are plotted. As only one SWV scan is needed in each measurement, the sample acquisition time can be less than 0.25 seconds at a 400-Hz SWV frequency.

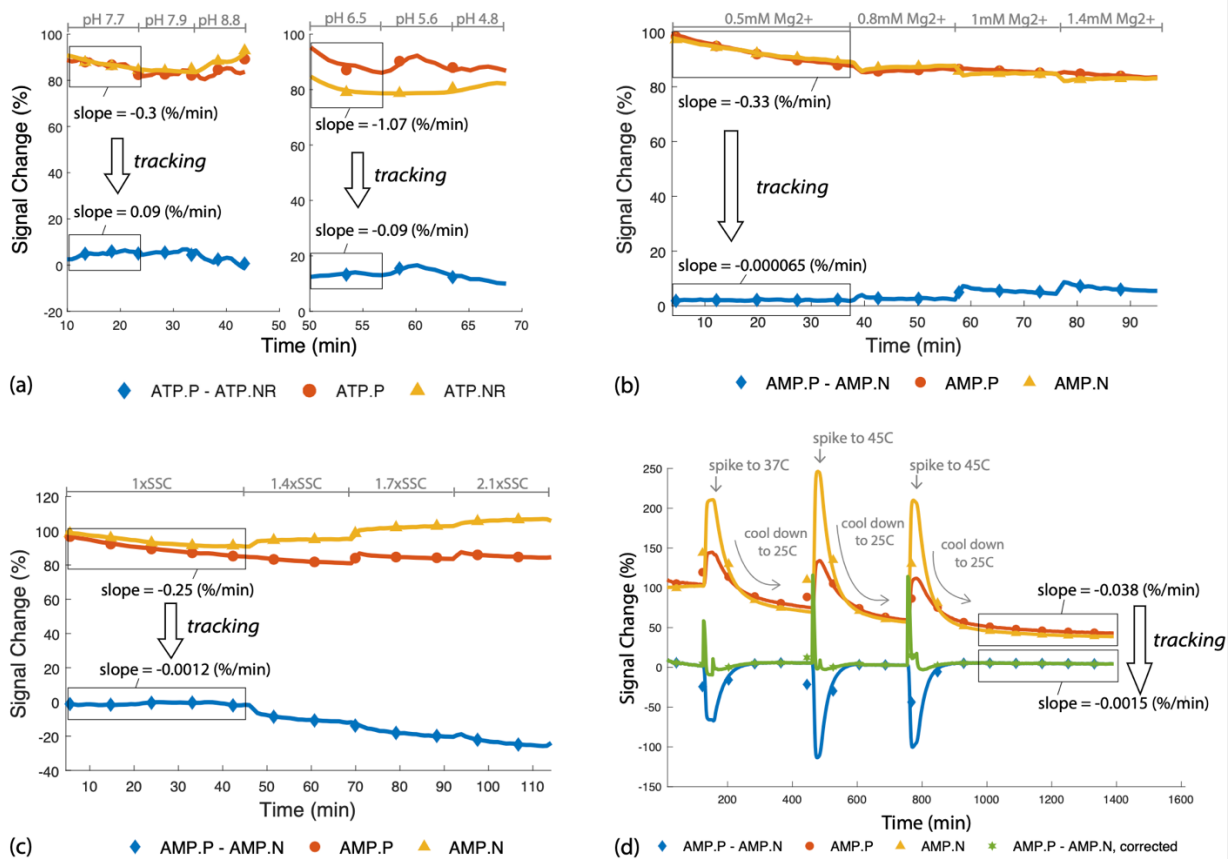

Figure 5. We study the responses from the aptamer pairs by modulating the environmental factors, including pH, ion concentration, and temperature. (a) The measurement results from the ATP.P/ATP.NR pair with pH varying between 4.8 and pH 8.8. (b) The measurement results from the AMP.P/AMP.N pair with $Mg^{2+}$ concentration varying from 0.5 to 1.4 mM. (c) The measurement results from the AMP.P/AMP.N pair with the ionic strength varying from 1× to 2.1×. (d) The measurement results from the AMP.P/AMP.N pair with varying temperatures. It is observed that these impacts cannot be fully canceled due to differences in temperature sensitivity. To compensate for this, we first model the temperature sensitivity of each aptamer using $2^{nd}$-order polynomial equations. Next, we correct the signals in each channel based on the estimated temperature profile. Last, we apply drift cancellation, and the result is shown in the green curve. As can be seen, the temperature impact can be mostly removed. The spikes are the artifacts due to insufficient temperature samples.

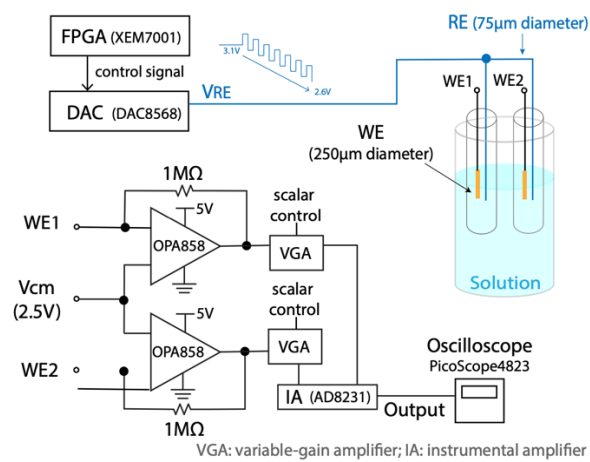

Figure 6. A two-channel pseudo-different current readout electronics is designed to double the throughput in dual-aptamer sensing by simultaneously measuring both aptamers. The SWV waveforms are generated from a digital-to-analog converter (DAC8568, Texas Instrument) using an FPGA (XEM7001, Opal Kelly Inc.). The DAC output drives a silver/silver chloride reference electrode (RE), and the two working electrodes are connected to the transimpedance amplifier (TIA) (OPA858, Texas Instrument) for signal amplification. The differential modes in the two outputs are extracted using an instrumentation amplifier (IA, AD8231, Analog Device Incs.). The final output is digitized using a high-resolution analog-to-digital converter (ADC). Here we use a multi-channel oscilloscope (PicoScope 4823). The data is analyzed in MATLAB.

Using aptamer pairs with differential signaling enables drift tracking and cancellation at an enhanced throughput, enabling long-term *in vivo* monitoring at an improved accuracy. The proposed technique is beneficial for structure-switching aptamers where KDM is not applicable. The main challenge lies in formulating aptamer pairs that could respond differently to the same target yet exhibit matched drift rates in the biofluids. Here we identify these pairs by (1) screening the existing aptamer pools and (2) modulating conjugating positions of the redox reporter. In the former, we screen the aptamers selected for small-molecule drugs [54] and other candidate sequences from the SELEX process [46]. We identify that the doxorubicin aptamer can pair with the ampicillin one due to their differential responses versus the ampicillin drug. We also find that both aptamers can serve as a pair for sensing doxorubicin. On the other hand, changing the redox reporter location to promote differential signaling is based on prior knowledge of the ATP aptamer conformation and the ISD construct in [6]. This is much more challenging to achieve, but our measurements show the pair indeed offers a better match in the individual drift rates due to the use of identical sequences. We believe there are many other approaches (or a combination of those), e.g., mixing with non-responsive aptamers or different sequence truncation schemes, that can further facilitate the selection of the paired aptamers.

One limitation of the screening selection approach is the potential interference from other responsive molecules. For example, our ampicillin aptamer pair would have ambiguous results if both the ampicillin and doxorubicin drugs were present in the sample simultaneously. Generally, this can be avoided as doxorubicin is primarily used in cancer therapy yet the antitumor effect of ampicillin is still unclear [28], and the possibility of administrating both drugs together is relatively low. In any case, one should cross-check any possible interference when formulating the aptamer pairs.

## *Conclusion*

This paper presents a dual aptamer approach for combating the drifts in E-AB biosensors. We discuss the rationale behind selecting these aptamer pairs and demonstrate effective cancellation of the sensor drifts in undiluted gold serum. We also study the influence of the pH, ion concentrations, and temperature and highlight that multi-parametric sensor fusion is necessary for mitigating their impacts. The dual aptamer scheme also offers throughput enhancement with custom-designed dual-channel readout electronics. The proposed approach can complement the existing KDM technique and facilitate the development of E-AB sensors for future clinical applications.

## *Acknowledgment*

The authors thank Yushan Young Scholar Program from Taiwan Ministry of Education (NTU-110VV001 and NTU- 111VV001), Taiwan National Science and Technology Council 110-2222-E-002-010-MY3, and NTU SoC Center.

## *Reference*

[1] Flores, M.; Glusman G.; Brogaard, K.;Price, N. D.; Hood, L. P4 medicine: How systems medicine will transform the healthcare sector and society. *Per. Med.* **2013**, 10, 565–576.

[2] Parolo, C.; Idili, A.; Ortega, G.; Csordas, A.; Hsu, A.; Arroyo-Curraś, N.; Yang, Q.; Ferguson, B. S.; Wang, J.; Plaxco, K. W. Real-Time Monitoring of a Protein Biomarker. *ACS Sens.* **2020**, 5, 1877−1881.

[3] Ferguson, B. S.; Hoggarth, D. A.; Maliniak, D.; Ploense, K.; White, R. J.; Woodward, N.; Hsieh, K.; Bonham, A. J.; Eisenstein, M.; Kippin, T.; Plaxco, K. W.; Soh, H. T. Real-time, aptamer-based tracking of circulating therapeutic agents in living animals. *Sci. Transl. Med*. **2013**, 5, 213ra165.

[4] Arroyo-Currás, N.; Somerson, J.; Vieira, P. A.; Ploense, K. L.; Kippin, T. E.; Plaxco, K. W. Real-time measurement of small molecules directly in awake, ambulatory animals. *Proc. Natl. Acad. Sci*. **2017**, 114, 645–650.

[5] Mage, P. L.; Ferguson, B. S.; Maliniak, D.; Ploense, K. L.; Kippin, T. E.; Soh, H. T. Closed-loop control of circulating drug levels in live animals. *Nat. Biomed. Eng*. **2017**, 1, 1–10.

[6] Wilson, B. D.; Hariri, A. A.; Thompson, I. A. P.; Eisenstein, M.; Soh, H. T. Independent control of the thermodynamic and kinetic properties of aptamer switches," *Nat. Commun* **2019**, 10, 1–9.

[7] Idili, A.; Arroyo-Curraś, N.; Ploense, K. L.; Csordas, A. T.; Kuwahara, M.; Kippin, T. E.; Plaxco, K. W. Seconds-resolved pharmacokinetic measurements of the chemotherapeutic irinotecan: In situ in the living body. *Chem. Sci.* **2019**, 10, 8164-8170.

[8] Dauphin-Ducharme, P.; Yang, K.; Arroyo-Currás, N.; Ploense, K. L.; Zhang, Y.; Gerson, J.; Kurnik, M.; Kappin, T. E.; Stojanovic, M. N.; Plaxco, K. W. Electrochemical Aptamer-Based Sensors for Improved Therapeutic Drug Monitoring and High-Precision, Feedback-Controlled Drug Delivery. *ACS Sens.* **2019**, 4, 10, 2832–2837.

[9] Arroyo-Currás, N.; Ortega, G.; Copp, D. A.; Ploense, K. L.; Plaxco, Z. A.; Kippin, T. E.; Hespanha, J. P.; Plaxco, K. W. High-Precision Control of Plasma Drug Levels Using Feedback-Controlled Dosing. *ACS Pharmacol. Transl. Sci.* **2018**, 1, 2, 110–118.

[10] Bock, L.; Griffin, L.; Latham, J.; Vermaas, E. H.; Toole, J. J. Selection of single-stranded DNA molecules that bind and inhibit human thrombin. *Nature* **1992**, 355, 564–566.

[11] Ellington, A.; Szostak, J. In vitro selection of RNA molecules that bind specific ligands. *Nature* **1990**, 346, 818–822.

[12] Tuerk, C.; Gold, L. Systematic evolution of ligands by exponential enrichment: RNA ligands to bacteriophage T4 DNA polymerase. *Science* **1990**, 249, 505–510.

[13] Leung, K. K.; Downs, A. M.; Ortega, G.; Kurnik, M.; Plaxco, K.W. Elucidating the Mechanisms Underlying the Signal Drift of Electrochemical Aptamer-Based Sensors in Whole Blood. *ACS Sens.* **2021**, 6, 3340−3347.

[14] Xu, X.; Makaraviciute, A.; Kumar, S.; Wen, C.; Sjödin, M.; Abdurakhmanov, E.; Danielson, U. H.; Nyholm, L.; Zhang, Z. Structural Changes of Mercaptohexanol Self-Assembled Monolayers on Gold and Their Influence on Impedimetric Aptamer Sensors. *Anal. Chem.* **2019**, 91, 14697−14704.

[15] Li, H.; Dauphin-Ducharme, P.; Ortega, G.; Plaxco, K. W. Calibration-Free Electrochemical Biosensors Supporting Accurate Molecular Measurements Directly in Undiluted Whole Blood. *J. Am. Chem. Soc.* **2017**, 139, 11207−11213.

[16] Li, H.; Arroyo-Curraś, N.; Kang, D.; Ricci, F.; Plaxco, K. W. Dual-Reporter Drift Correction to Enhance the Performance of Electrochemical Aptamer-Based Sensors in Whole Blood. *J. Am. Chem. Soc.* **2016**, 138, 15809−15812.

[17] Downs, A. M.; Gerson, J.; Leung, K. K.; Honeywell, K. M.; Kippin, T.; Plaxco, K. W. Improved calibration of electrochemical aptamer-based sensors. *Nature* **2022**, 12, 5535.

[18] Tuerk, C.; Gold, L. Systematic Evolution of Ligands by Exponential Enrichment: RNA Ligands to Bacteriophage T4 DNA Polymerase. *Science* **1990**, 249, 505–510.

[19] Robertson, D. L.; Joyce, G. F. Selection *in vitro* of an RNA enzyme that specifically cleaves single-stranded DNA. *Nature* **1990**, 344, 467–468.

[20] Ellington, A.D.; Szostak, J.W. *In vitro* selection of RNA molecules that bind specific ligands. *Nature* **1990**, 346, 818–822.

[21] Greiner, J. V.; Glonek, T. Intracellular ATP Concentration and Implication for Cellular Evolution. *Biology* **2021**, 10, 1166.

[22] Giachetto, G.; Pirez, M. C.; Nanni, L. P.; Martínez, A.; Montano, A.; Algorta, G.; Kaplan, S. L.; Ferrari, A. Ampicillin and Penicillin Concentration in Serum and Pleural Fluid of Hospitalized Children With Community-Acquired Pneumonia. *The Pediatric Infectious Disease Journal* **2004**, 625-629.

[23] Li, H.; Dauphin-Ducharme, P.; Arroyo-Currás, N.; Tran, C. H.; Vieira, P. A.; Li, S.; Shin, C.; Somerson, J.; Kippin, T. E.; Plaxco, K. W. A Biomimetic Phosphatidylcholine-Terminated Monolayer Greatly Improves the In Vivo Performance of Electrochemical Aptamer-Based Sensors. *Angew Chem Int Ed Engl.* **2017**, 56, 7492–7495.

[24] Chien, J. C.; Baker, S. W.; Soh, H. T. Design and Analysis of a Sample-and-Hold CMOS Electrochemical Sensor for Aptamer-based Therapeutic Drug Monitoring. *IEEE JSSC*, **2020**, 55, 2914-2929.

[25] Arroyo-Curraś, N.; Dauphin-Ducharme, P.; Ortega, G.; Ploense, K. L.; Kippin, T. E.; Plaxco, K. W. Subsecond-Resolved Molecular Measurements in the Living Body Using Chronoamperometrically Interrogated Aptamer-Based Sensors. *ACS Sens.* **2018**, 3, 360−366.

[26] Downs, A. M.; Gerson, J.; Ploense, K. L.; Plaxco, K. W.; Dauphin-Ducharme, P. Subsecond-Resolved Molecular Measurements Using Electrochemical Phase Interrogation of Aptamer-Based Sensors. *Anal. Chem.* **2020**, 92, 14063−14068.

[27] Kang, D.; Ricci, F.; White, R. J.; Plaxco, K. W. Survey of Redox-Active Moieties for Application in Multiplexed Electrochemical Biosensors. *Anal. Chem.* **2016**, 88, 21, 10452–10458.

[28] Hut, E. F.; Radulescu, M.; Pilut, N.; Macasoi, I.; Berceanu, D.; Coricovac, D.; Pinzaru, I.; Cretu, O.; Dehelean, C. Two Antibiotics, Ampicillin and Tetracycline, Exert Different Effects in HT-29 Colorectal Adenocarcinoma Cells in Terms of Cell Viability and Migration Capacity. *Curr Oncol.* **2021**, 28, 2466–2480.

[29] Velicer, C. M.; Heckbert, S. R.; Lampe, J. W.; Potter, J. D.; Robertson, C. A.; Taplin, S. H. Antibiotic Use in Relation to the Risk of Breast Cancer. *JAMA*. **2004**, 291, 827-835.

[30] Lia, L.; Lia, X.; Zhong, W.; Yang, M.; Xu, M.; Sun, Y.; Ma, J.; Liu, T.; Song, X.; Dong, W.; Liu, X.; Chen, Y.; Liu, Y.; Abla, Z.; Liu, W.; Wang, B.; Jiang, K.; Cao, H. Gut microbiota from colorectal cancer patients enhances the progression of intestinal adenoma in *Apc*$^{min/+}$ mice. *EBioMedicine* **2019**, 48, 301–315.

[31] Jeong S.; Paeng, I. R. Sensitivity and Selectivity on Aptamer-Based Assay: The Determination of Tetracycline Residue in Bovine Milk. *The Scientific World Journal* **2012**, 159456.

[32] Roy, S.; Hennelly, S. P.; Lammert, H.; Onuchic, J. N.; Sanbonmatsu, K. Y. Magnesium controls aptamer-expression platform switching in the SAM-I riboswitch. *Nucleic Acids Research*, **2019**, 47, 3158-3170.

[33] Yu, Z.; Lai, R. Y. A reagentless and reusable electrochemical aptamer-based sensor for rapid detection of ampicillin in complex samples. *Talanta* **2018**, 176, 619–624.

[34] Yu, Z.; Sutlief, A. L.; Lai, R. Y. Towards the development of a sensitive and selective electrochemical aptamer-based ampicillin sensor. *Sensors and Actuators B* **2018**, 258, 722–729.

[35] Agteresch, H. J.; Dagnelie, P. C.; van den Berg, J. W. O.; Wilson, J. H. P. Adenosine Triphosphate: Established and Potential Clinical Applications. *Drugs* **1999**, 58, 211–232.

[36] Carlton, K. K.; Lee, P. D.; MPH. Drug Dosages. Harriet Lane Handbook Twenty Second Edition **2021**, Chapter 30, 665-1076

[37] Dan, C.; Chien, J. C.; Axpe, E.; Blankemeier, L.; Baker, S. W.; Swaminathan, S.; Piunova, V. A.; Zubarev, D. Y.; Maikawa, C. L.; Grosskopf, A. K.; Mann, J. L.; Soh, H. T.; Appel, E. A. Combinatorial Polyacrylamide Hydrogels for Preventing Biofouling on Implantable Biosensors. *Adv. Mater.* **2022**, 34, 2109764.

[38] Chrusch, C.; Bautista, E.; Jacobs, H. K.; Light, R. B.; Bose, D.; Duke, K.; Mink, S. N. Blood pH Level Modulates Organ Metabolism of Lactate in Septic Shock in Dogs. *Journal of Critical Care* **2022**, 17, 188-202.

[39] Drewry, A. M.; Fuller, B. M.; Bailey, T. C.; Hotchkiss, R. S. Body temperature patterns as a predictor of hospital-acquired sepsis in afebrile adult intensive care unit patients: a case-control study. *Critical Care* **2013**, 17, R200.

[40] Razmjou, A.; Asadnia, M.; Hosseini, E.; Korayem, A. H.; Chen, V. Design principles of ion selective nanostructured membranes for the extraction of lithium ions. *Nat Commun* **2019**, 10, 5793.

[41] Chin, Y. L.; Chou, J. C.; Sun, T. P.; Chung, W. Y.; Hsiung, S. K. A novel pH sensitive ISFET with on chip temperature sensing using CMOS standard process. *Sensors and Actuators* **2001**, 72, 582-593.

[42] Thompson, I.A.P.; Zheng, L.; Eisenstein, M.; Soh, H. T. Rational design of aptamer switches with programmable pH response. *Nat Commun* **2020**, 11, 2946.

[43] Amodio, A.; Zhao, B.; Porchetta, A.; Idili, A.; Castronovo, M.; Fan, C.; Ricci, F. Rational Design of pH-Controlled DNA Strand Displacement. *J. Am. Chem. Soc.* **2014**, 136, 16469−16472.

[44] Li, L.; Jiang, Y.; Cui, C.; Yang, Y.; Zhang, P.; Stewart, K.; Pan, X.; Li, X.; Yang, L.; Qiu, L.; Tan, W. Modulating Aptamer Specificity with pH-Responsive DNA Bonds. *J. Am. Chem. Soc.* **2018**, 140, 41, 13335–13339.

[45] Li, S.; Ferrer-Ruiz, A.; Dai, J.; Ramos-Soriano, J.; Du, X.; Zhu, M.; Zhang, W.; Wang, Y.; Herranz, M. A.; Jing, L.; Zhang, Z.; Li, H.; Xia F.; Martín, N. A pH-independent electrochemical aptamer-based biosensor supports quantitative, real-time measurement in vivo. *Chem. Sci.*, **2022**, 13, 8813.

[46] Huang, C. J.; Lin, H. I.; Shiesh, S. C.; Lee, G. B. Integrated microfluidic system for rapid screening of CRP aptamers utilizing systematic evolution of ligands by exponential enrichment (SELEX). *Biosensors and Bioelectronics* **2010**, 25, 1761-1766.

[47] Phares, N.; White, R. J.; Plaxco, K. W. Improving the Stability and Sensing of Electrochemical Biosensors by Employing Trithiol-Anchoring Groups in a Six-Carbon Self-Assembled Monolayer. *Anal. Chem.* **2009**, 81, 1095–1100.

[48] Barberis, A,; Petrini, E. M.; Mozrzymas, J. W. Impact of synaptic neurotransmitter concentration time course on the kinetics and pharmacological modulation of inhibitory synaptic currents. *Front Cell Neurosci.* **2011,** 5:6.

[49] Mahlum, J. D.; Pellitero, M. A.; Arroyo-Currás, N. Chemical Equilibrium-Based Mechanism for the Electrochemical Reduction of DNA-Bound Methylene Blue Explains Double Redox Waves in Voltammetry. *The Journal of Physical Chemistry* **2021**, 125 (17), 9038-9049.

[50] Nakatsuka, N.; Yang, K. A.; Abendroth, J. M.; Cheung, K. M.; Xu, X.; Yang, H.; Zhao, C.; Zhu, B.; Rim, Y. S.; Yang, Y.; Weiss, P. S.; Stojanović, M. N.; Andrews, A. M. Aptamer–field-effect transistors overcome Debye length limitationsfor small-molecule sensing. *Science* **2018**, 362, 319–324.

[51] Somerson, J.; Plaxco, K. W. Electrochemical Aptamer-Based Sensors for Rapid Point-of-Use Monitoring of the Mycotoxin Ochratoxin A Directly in a Food Stream. *Molecules.* **2018**, 23(4), 912.

[52] Huang, Y. J.; Tzeng, T. H.; Lin, T. W.; Huang, C. H.; Yen, P. W.; Kuo, P. H.; Lin, C. T.; Lu, S. S. A Self-Powered CMOS Reconfigurable Multi-Sensor SoC for Biomedical Applications. *IEEE Journal of Solid-State Circuits* **2014**, 49, 851-866.

[53] Cacho-Soblechero, M. Malpartida-Cardenas, K.; Cicatiello, C.; Rodriguez-Manzano, J.; Georgiou, P. A Dual-Sensing Thermo-Chemical ISFET Array for DNA-Based Diagnostics. *IEEE Transactions on Biomedical Circuits and Systems* **2020**, 14, 477-489.

[54] Mehlhorn, A.; Rahimi, P.; Joseph, Y. Aptamer-Based Biosensors for Antibiotic Detection: A Review. *Biosensors* **2018**, 8, 54.

# Dual-aptamer Drift Cancelling Techniques to Improve Long-term Stability of Real-Time Structure-Switching Aptasensors

Ya-Chen Tsai[1], Wei-Yang Weng[2], Yu-Tong Yeh[1], Jun-Chau Chien[1,2]

[1]Department of Electrical Engineering

[2]Graduate Institute of Electronics Engineering

National Taiwan University

## *Table of content*

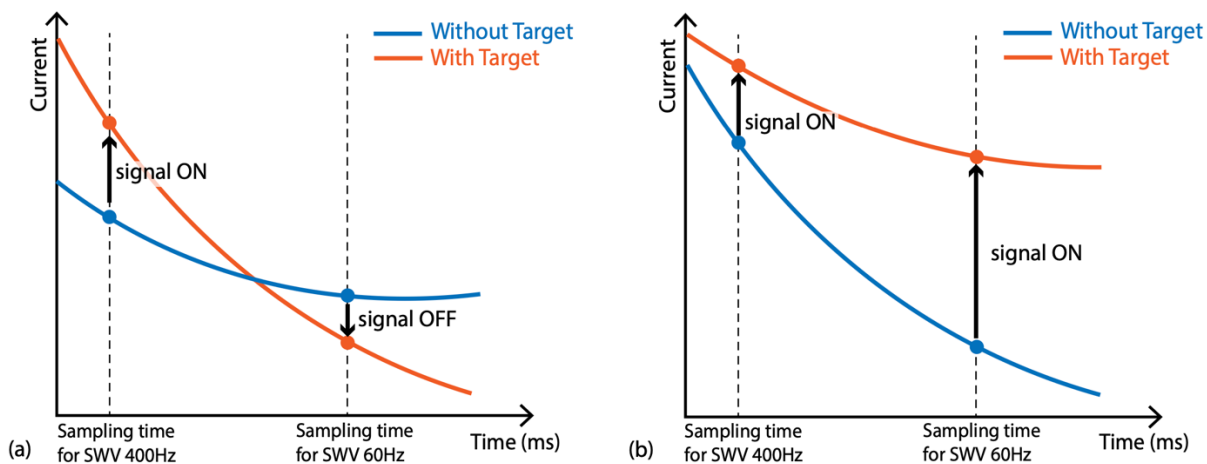

Figure S1: (a) An illustration of the kinetic differential measurement (KDM) technique. When the electron transfer (ET) kinetics of the redox reporter is modulated by the presence of the target, E-AB sensors can exhibit opposite gain polarity when measured at high and low SWV frequencies. This is due to the differences in their sampling instants. By subtracting one signal from the other, drifts are removed while the signals are boosted. (b) We believe that some aptamers can have modulation in both the ET kinetics and the absolute current magnitude upon target binding. This leads to identical gain polarity, and KDM is no longer applicable.

| Sequence name | Target | Sequence | Ref |
|---|---|---|---|
| AMP.P | Ampicillin | 5'-SS-(CH2)6-TTAGTTGGGGTTCAGTTG-G-(CH2)7-NH-MB-3' | 33, 34 |
| AMP.N | | 5'-SS-(CH2)6-ACCATCTGTGTAAGGGGTAAGGGGTGGT-(CH2)7-NH-MB-3' | 3 |
| ATP.P | ATP | 5'-SS-(CH2)6-CACCTGGGGGAGTATTGCGGAGGAAGGT(MB)TTTTCCAGGTG-3' | 6 |
| ATP.NR | | 5'-SS-(CH2)6-CACCTGGGGGAGTATTGCGGAGGAAGGTTTTTCCAGGTG-(CH2)7-NH-MB-3' | 6 |

Table S1. The four sequences used in this work. The 'P' indicates a positive (signal-on) response, the 'N' indicates a negative (signal-off) response, and the 'NR' stands for non-responding. Each sequence has one disulfide bond at the 5' terminus followed by a six-carbon thiol linker. Redox reporters are labeled at the 3' terminus in the AMP.P, AMP.N, and ATP.NR sequences but at the middle of the ATP.P sequences.

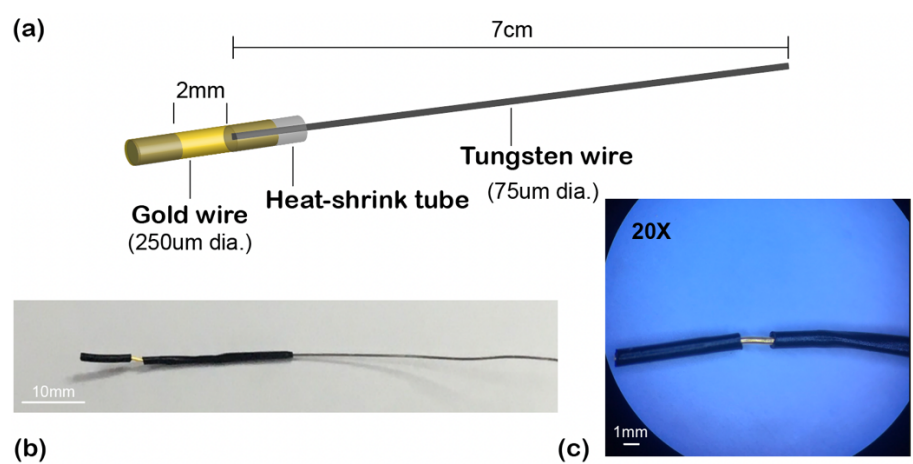

Figure S2. (a) The schematic of our sensing device. The aptamers are immobilized on the surface of a gold wire at a diameter of 250μm. The sensing area is defined manually using black heat-shrink tubing, which also serves as a spacer to protect the aptamers from being scratched during the insertion into the circulation system. We employ a similar strategy in preparing the counter (platinum) and reference (Ag/AgCl) electrodes. (b) and (c) Device photos.

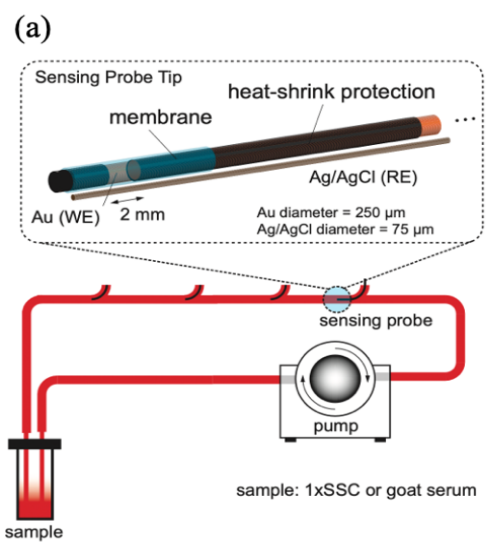 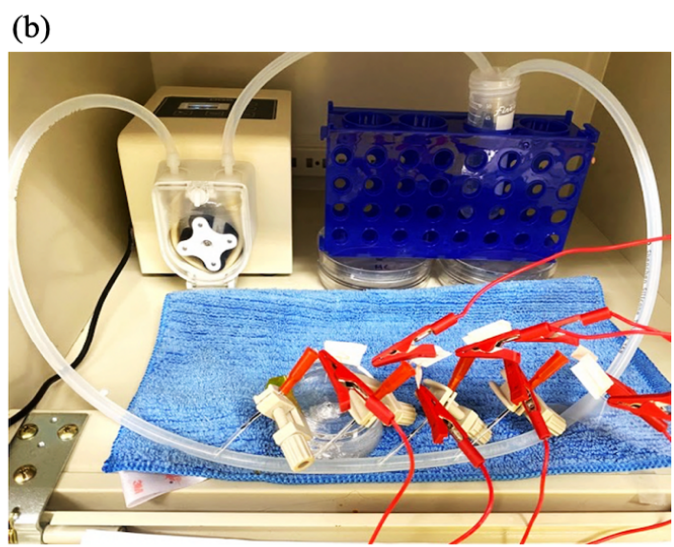

Figure S3. (a) The schematic and (b) setup photo of the circulation system for the long-term studies of aptamer signaling to simulate the situation in the implantable system. Multiple catheters are used to enable study replication. All devices except the potentiostat are placed in a Faraday cage to minimize environmental noise.

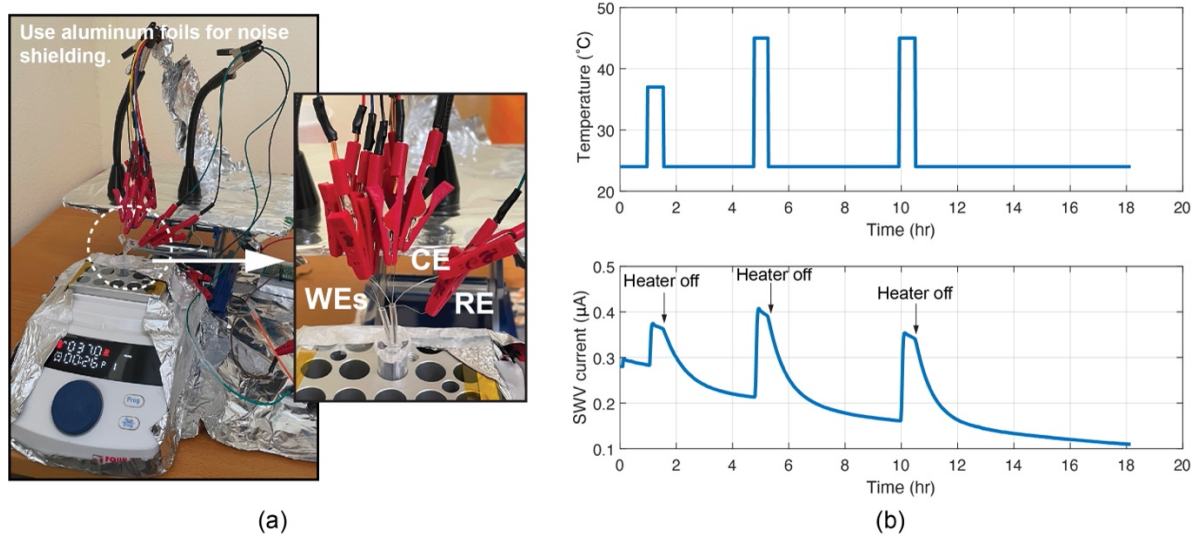

Figure S4: (a) The testing setup for studying the temperature effects on the aptamer pairs. Wire probes are inserted into the buffer through tygon tubings to avoid wires sticking to the sidewalls. Eight probes are prepared for each study. (b) Top: the temperature profile from the thermal controller. The aptamers are first monitored at room temperature for ~1hr. Next, the temperature is elevated to 37°C for 30 minutes. The heater is then turned off to allow the aluminum block cool to room temperature passively. This explains the exponential decaying behavior observed in the measured SWV signal (bottom). The process is repeated twice at an elevated temperature of 45°C.

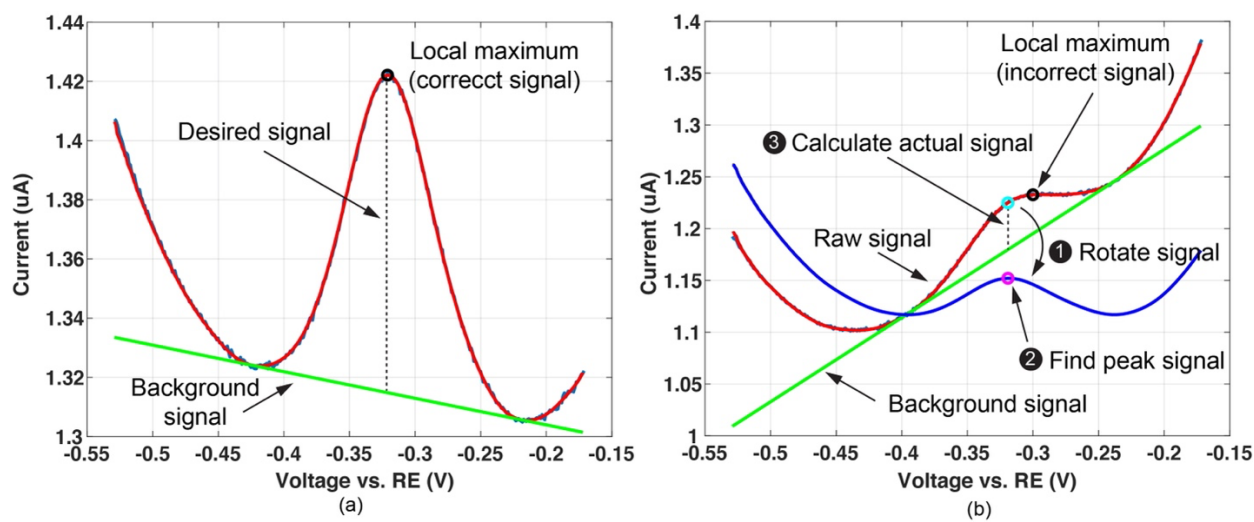

Figure S5: Data processing flow for extracting signals from square-wave voltammograms with tilted baseline. Generally, we find the SWV signal by subtracting the peak current from a baseline signal tangential to the two nearby local minimums (Fig. S4(a)). However, the signal peak reduces and the baseline tilted significantly after >12 hr of continuous recording, leading to incorrect localization of the peak signal (Fig. S4(b)). We remedy this issue by first rotating the raw SWV waveform, finding the potential where the peak signal locates, and subtracting the peak current in the raw voltammogram from its corresponding background.

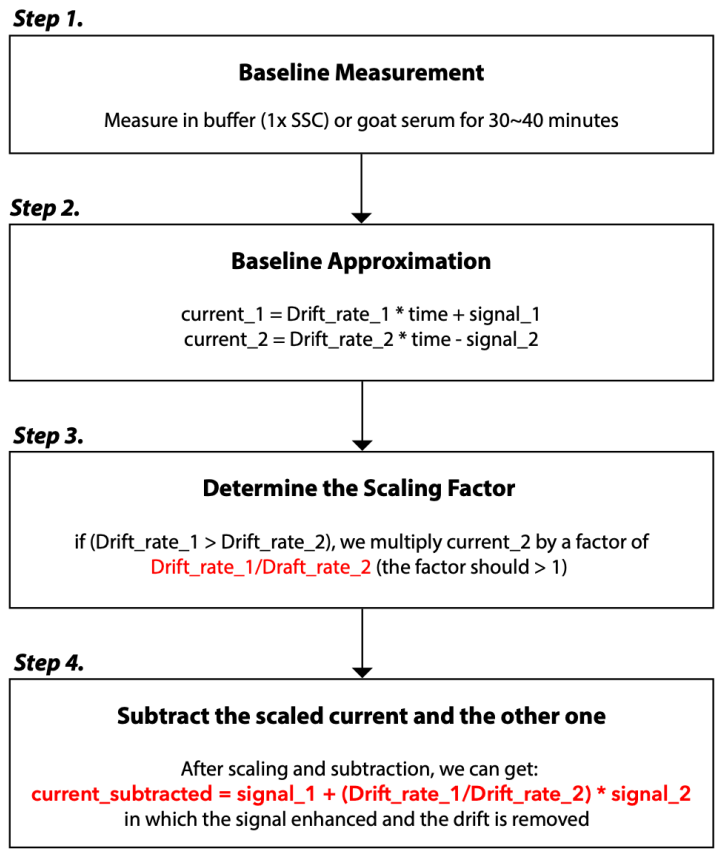

Figure S6. Drift-canceling signal-processing flowchart. First, we record the SWV signals continuously in buffer (1x SSC) or goat serum (without target molecules) for 10+ minutes. Next, we scale the signal from one of the aptamer channels and perform signal subtraction to remove the drifts.

**Mathematical Model.** A mathematical model is provided that can accurately predict the dissociation constants and Langmuir coefficients after drift removal. From the Langmuir isotherm, the binding of aptamers and targets can be described as (1), here 'x' represents the target concentration, [Ax] represents the amounts of aptamer-target compounds, $A_0$ is the Langmuir coefficient, and $K_D$ is the dissociation constant:

$$[AX] = \frac{A_0[x]}{[x] + K_D} \quad (1)$$

We can then describe the signals from the positive strands (P) and the negative strands (N) as (2) and (3) respectively:

$$[PX] = \frac{P_0[x]}{[x] + K_{Dp}} + drift\_1 \quad (2)$$

$$-[NX] = \frac{-N_0[x]}{[x] + K_{Dn}} + drift\_2 \quad (3)$$

Here the Langmuir coefficients have included the scaling, therefore we assume the two '*drift*' are the same in (2) and (3). Note that we assume $P_0$ and $N_0$ are positive values, and we added a minus to represent the negative response. Next, we subtracted (2) and (3) and approximated the result as another Langmuir isotherm as (4):

$$signal = [PX] + [NX] = \frac{[x]\{(P_0 + N_0)[x] + (P_0 K_{Dn} + N_0 K_{Dp})\}}{[x]^2 + (K_{Dp} + K_{Dn})[x] + K_{Dp} K_{Dn}} = \frac{S_0[x]}{[x] + K_D} \quad (4)$$

Based on the knowledge of the dissociation – when [x] = $K_D$, the signal should equal half of the saturation value ($P_0 + N_0$) – we derived the Langmuir coefficient after subtraction $S_0$ as the sum of the Langmuir coefficients of the two strands, and the dissociation constant $K_D$ as (5):

$$K_D = \frac{(P_0 - N_0)(K_{Dp} - K_{Dn}) + \sqrt{(P_0 - N_0)^2(K_{Dp} + K_{Dn})^2 + 16 K_{Dp} K_{Dn} P_0 N_0}}{2(P_0 + N_0)} \quad (5)$$

Particularly if $P_0 = N_0$, the dissociation constant after subtraction is equal to the geometric mean of original $K_D$s (6):

$$K_D = \sqrt{K_{Dp}K_{Dn}} \quad (6)$$

That is, we can divide the signals by the Langmuir coefficients before subtraction and get the new dissociation constant by an easy calculation. We got Langmuir coefficients and dissociation constants from experiments that were highly analogous to the predicted values, shown in Table S2.

(a)

| AMP.P AMP.N | $P_0$ | $N_0$ | $S_0$ | $S_0$ (mM) calculated | error | $K_{Dp}$ (mM) | $K_{Dn}$ (mM) | $K_D$ (mM) | $K_D$ (mM) calculated | error |
|---|---|---|---|---|---|---|---|---|---|---|
| 60hz | 0.38 | 0.24 | 0.623 | 0.62 | 0.5% | 2.38 | 2 | 2.25 | 2.225 | 1.1% |
| 100hz | 0.573 | 0.44 | 1.013 | 1.013 | 0% | 4.05 | 3.95 | 4.01 | 4.006 | 0.09% |
| 200hz | 0.22 | 0.286 | 0.502 | 0.506 | 0.8% | 1.83 | 2.34 | 2.05 | 2.1 | 2.6% |

(b)

| ATP.P ATP.NR | $P_0$ | $N_0$ | $S_0$ | $S_0$ (mM) calculated | error | $K_{Dp}$ (mM) | $K_{Dn}$ (mM) | $K_D$ (mM) | $K_D$ (mM) calculated | error |
|---|---|---|---|---|---|---|---|---|---|---|
| 60hz | 7.2 | 0.26 | 7.33 | 7.46 | 1.7% | 7.14 | 34.4 | 7.22 | 7.479 | 3.6% |
| 100hz | 8.7 | 0.069 | 8.78 | 8.769 | 0.1% | 7.45 | 1.5 | 7.4 | 7.372 | 0.37% |
| 200hz | 10.9 | 0.065 | 10.94 | 10.965 | 0.2% | 7.77 | 2.8 | 7.62 | 7.727 | 1.4% |

Table S2. The extracted Langmuir coefficients from the experiments. (a) The coefficients for the AMP.P/AMP.N pair. (b) The coefficients for the ATP.P/ATP.NR pair. Studies are performed at an SWV frequency of 60 Hz, 100 Hz, and 200 Hz, respectively. As the ATP.P has a much larger sensitivity than the ATP.NR aptamer, the composite $K_D$ is mainly determined by that of the ATP.P aptamer. On the other hand, both AMP.P and AMP.N aptamers exhibit equal but opposite responses. The composite $K_D$ is thus equal to their geometric mean. A close match between the prediction and the measurements is observed.

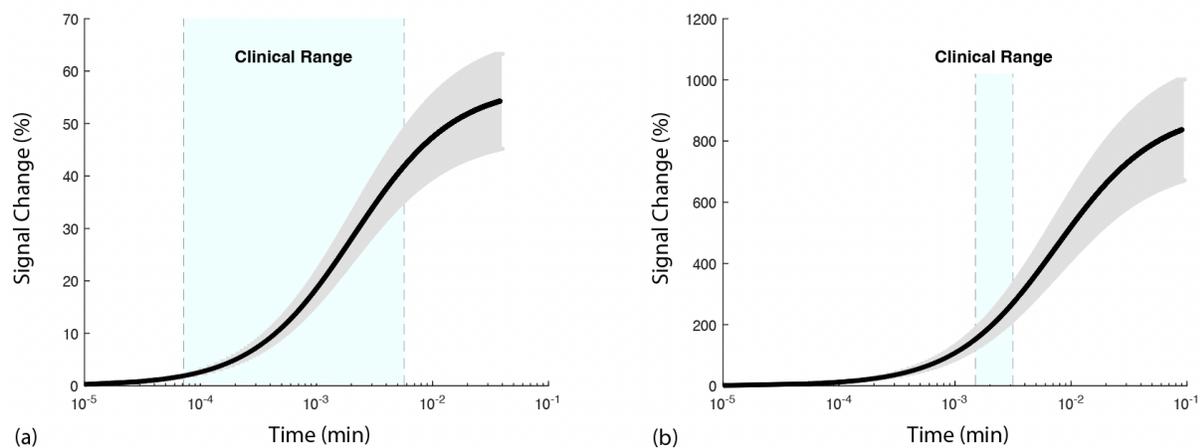

Figure S7. Langmuir curves for the (a) ampicillin aptamer pair (titrated to 7.6 mM) and (b) ATP aptamer pair (titrated to 18 mM). The black curves are the averaged Langmuir approximation from three experiments (after drift cancellation). The grey shadows are the distribution. The dissociation constant of the ampicillin aptamer pair is $3.19 \pm 0.93\ mM$ with a maximum deviation of 14.9% within the clinical range (71.5uM ~ 5.7mM [36]). The dissociation constant of the ATP aptamer pair is $7.41 \pm 0.916\ mM$ and a maximum deviation of 18% within the clinical range (1.5mM~3.15mM [21, 35]). All titration experiments were conducted in the buffer.

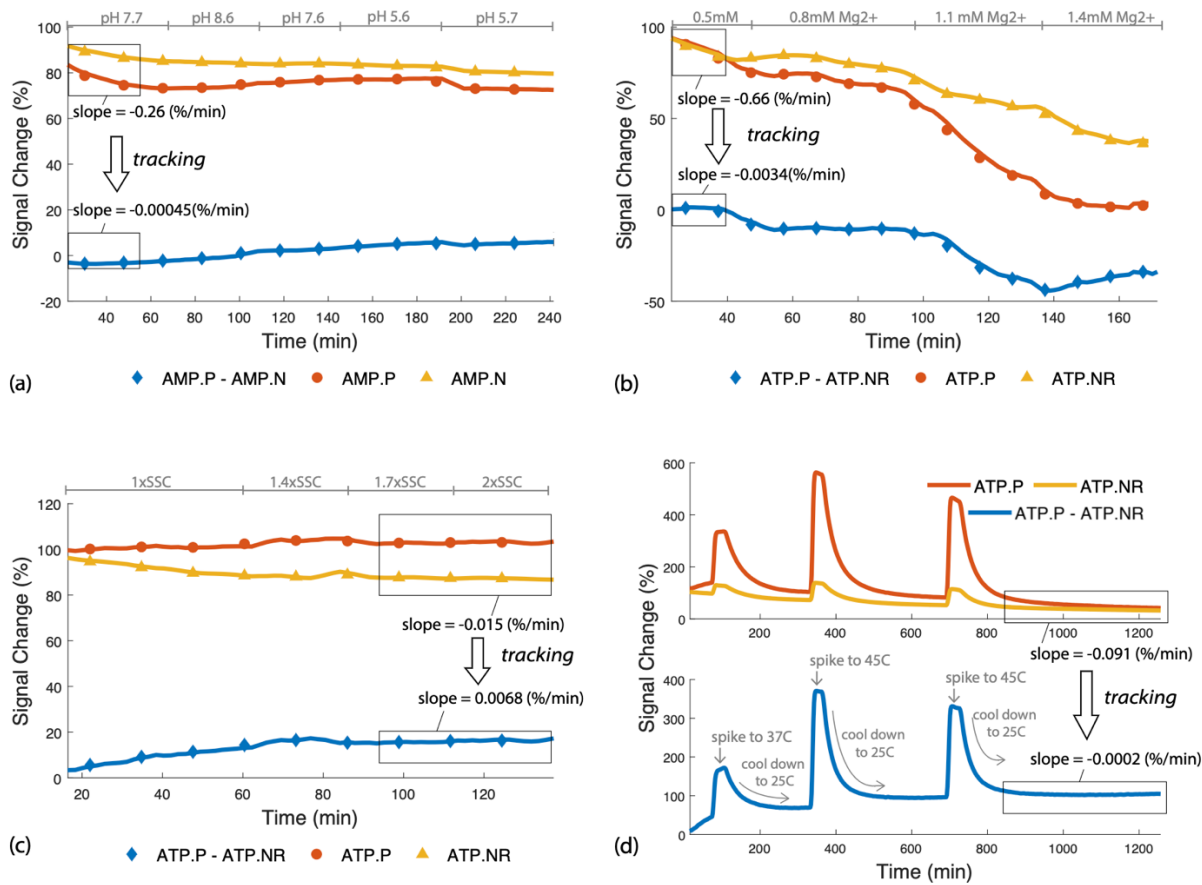

Figure S8. Additional measurement results from modulating the pH, ion concentration, and temperatures. (a) The results from modulating pH in the AMP.P/AMP.N aptamer pair. A 578-fold drift reduction is measured. (b) The results from modulating $Mg^{2+}$ concentration in the ATP.P/AMP.NR aptamer pair. A 178-fold drift reduction is measured. (c) The results from modulating ionic strength in the ATP.P/AMP.NR aptamer pair. A 2.2-fold drift reduction is measured. (d) The results from modulating the temperature in the ATP.P/AMP.NR aptamer pair. A 455-fold drift reduction is measured.

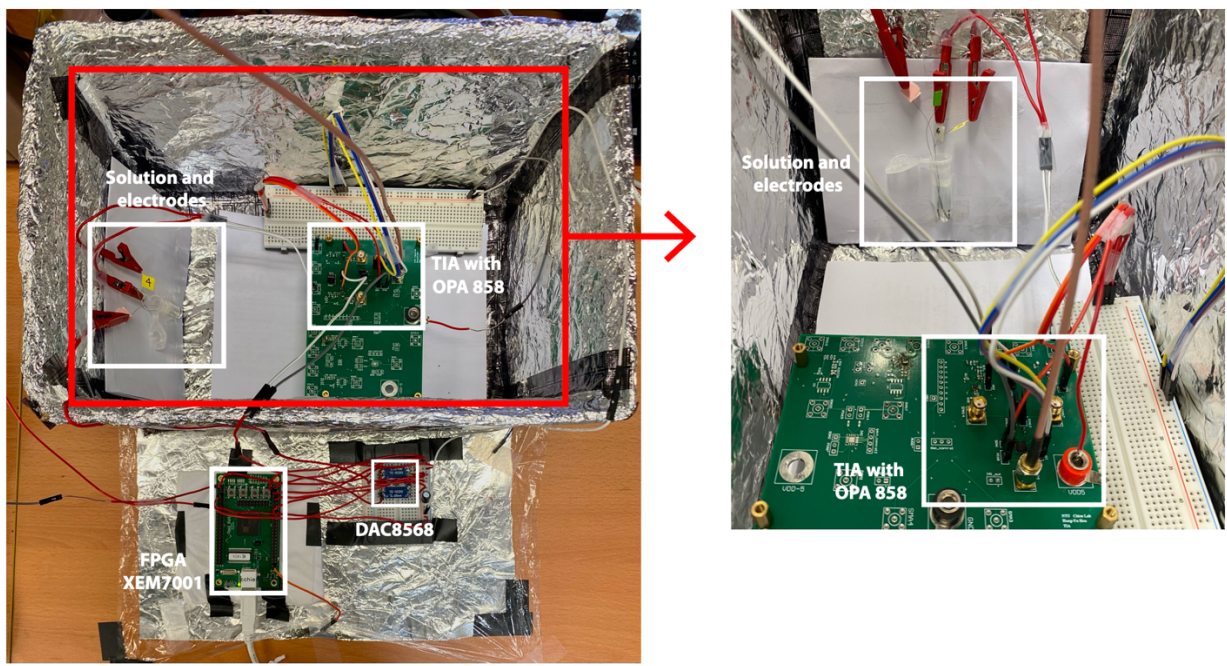

Figure S9. Our hardware setup photo. A two-electrode (RE-WE) electrochemical cell is used in our testing. To ensure high signal quality, we place the working electrode (WE) and the reference electrode (RE) within a very short distance. We use an FPGA (XEM7001, AMD Xilinx) to generate the square-wave voltammetry (SWV) waveforms. The two WEs are connected to the two transimpedance amplifiers (TIA) implemented with an operational amplifier (OPA858, Texas Instruments) and a 1-M$\Omega$ feedback resistor. This transforms the redox currents into voltages. The outputs of the TIA are connected to an instrumentation amplifier (IA) (AD8231, Analog Devices Inc.) for signal subtraction. An oscilloscope (PicoScope 4823, Pico Technology) is used to digitize the signals.